\documentclass[letterpaper,11pt,amsfonts,amssymb]{article}

\usepackage{graphics,graphicx,psfrag,color,float}
\usepackage{epsfig}
\usepackage{amsmath}
\usepackage{amssymb}

\begin{document}
\newcommand{\hrho}{\widehat{\rho}}
\newcommand{\hsigma}{\widehat{\sigma}}
\newcommand{\homega}{\widehat{\omega}}
\newcommand{\hI}{\widehat{I}}
\newcommand*{\spr}[2]{\langle #1 | #2 \rangle}
\newcommand*{\bbN}{\mathbb{N}}
\newcommand*{\bbR}{\mathbb{R}}
\newcommand*{\cB}{\mathcal{B}}
\newcommand*{\barpi}{\overline{\pi}}
\newcommand*{\barP}{\overline{P}}
\newcommand*{\eps}{\varepsilon}
\newcommand*{\id}{I}
\newcommand{\orho}{\overline{\rho}}
\newcommand{\omu}{\overline{\mu}}
\newcommand*{\half}{{\frac{1}{2}}}
\newcommand*{\ket}[1]{| #1 \rangle}
\newcommand{\trho}{{\widetilde{\rho}}_n^\gamma}
\newcommand{\ttrho}{{\widetilde{\rho}}}
\newcommand{\tsigma}{{\widetilde{\sigma}}}
\newcommand{\tpi}{{\widetilde{\pi}}}
\newcommand*{\bra}[1]{\langle #1 |}
\newcommand*{\proj}[1]{\ket{#1}\bra{#1}}
\newcommand{\otrho}{{\widetilde{\rho}}_n^{\gamma 0}}
\newcommand{\be}{\begin{equation}}
\newcommand{\bea}{\begin{eqnarray}}
\newcommand{\eea}{\end{eqnarray}}
\newcommand{\tr}{\mathrm{Tr}}
\newcommand*{\Hmin}{H_{\min}}
\newcommand{\rank}{\mathrm{rank}}
\newcommand{\tends}{\rightarrow}
\newcommand{\uS}{\underline{S}}
\newcommand{\oS}{\overline{S}}
\newcommand{\uD}{\underline{D}}
\newcommand{\oD}{\overline{D}}
\newcommand{\ee}{\end{equation}}
\newcommand{\supp}{\rm{supp}}
\newcommand{\n}{{(n)}}
\newtheorem{definition}{Definition}
\newtheorem{theorem}{Theorem}
\newtheorem{proposition}{Proposition}
\newtheorem{lemma}{Lemma}
\newtheorem{defn}{Definition}
\newtheorem{corollary}{Corollary}
\newcommand{\qed}{\hspace*{\fill}\rule{2.5mm}{2.5mm}}
\newcommand{\beq}{\begin{equation}}
\newcommand{\enq}{\end{equation}}
\newcommand{\beqa}{\begin{eqnarray}}
\newcommand{\enqa}{\end{eqnarray}}
\newcommand{\beqan}{\begin{eqnarray*}}
\newcommand{\enqan}{\end{eqnarray*}}
\newcommand{\mycal}[1]{\mathcal #1}
\newcommand\figcaption{\def\@captype{figure}\caption}
\newcommand{\mbm}[1]{\pmb{#1}}

\newenvironment{proof}{\noindent{\it Proof}\hspace*{1ex}}{\qed\medskip}
\def\reff#1{(\ref{#1})}

\title{Extensions of the quantum Fano inequality}

\author{Naresh Sharma \\
Tata Institute of Fundamental Research \\
Mumbai 400 005, India \\
nsharma@tifr.res.in
}

\date{\today}

\maketitle

\begin{abstract}
Quantum Fano inequality (QFI) in quantum information theory provides an
upper bound to the entropy exchange by a function of the entanglement fidelity.
We give various Fano-like upper bounds to the entropy exchange and QFI is a special case
of these bounds. These bounds also give an alternate derivation of the QFI.
\end{abstract}

\section{Introduction}

Classical Fano inequality (CFI) in classical information theory provides an upper
bound to the conditional entropy of two correlated random variables say $X$ and $Y$.
Suppose we wish
to obtain an estimate of $X$ when $Y$ is known. To get an estimate of $X$, we
compute a function of $Y$, denoted by $\hat{X}$. Let $n$ be the cardinality of the
set from which $X$ takes values. CFI upper bounds the conditional Shannon
entropy of $X$ given $Y$, denoted by $H_S(X|Y)$,
by a function of the probability of success defined as
\beq
P_s = \Pr\{\hat{X}=X\}
\enq
(see p. 37 in \cite{covertom}) and is given by
\beq
H_S(X|Y) \leq H(P_s) + (1-P_s) \ln(n-1),
\enq
where
\beq
\label{binent}
H(x) = -x\ln(x) - (1-x)\ln(1-x)
\enq
is the binary entropy function.
CFI is useful in proving
the converse to the Shannon's noisy channel coding theorem
(see p. 206 in Ref. \cite{covertom}).

QFI provides an upper bound to the entropy exchange by
a function of the entanglement fidelity, and the function is similar to the
function of the probability of success used in the CFI.

More specifically,
let $R$ and $Q$ be two quantum systems described by a Hilbert space
$\mycal{H}_Q$ of finite dimension $d$, where $d \geq 2$. The
joint system $RQ$ is initially prepared in a pure entangled state
\beq
\ket{ \psi^{RQ} } = \sum_{k=1}^{d} \sqrt{\lambda_k} \ket{k^R} \ket{k^Q},
\enq
where
$\mbm{\lambda} = [\lambda_1 \cdots \lambda_d]$ is a probability vector, i.e.,
$\lambda_k \geq 0$, $\sum_{k=1}^d \lambda_k = 1$, $\{ \ket{k^R} \}$
and $\{\ket{k^Q}\}$, $k=1,...,d$, are two orthonormal bases for
$\mycal{H}_Q$. $\ket{ \psi^{RQ} }$
is a purification of $\rho$, the state of system $Q$, and
\beq
\label{eqrho}
\rho = \tr_R(\ket{ \psi^{RQ} } \bra{ \psi^{RQ} }) = \sum_{k=1}^d \lambda_k
\ket{k^Q} \bra{k^Q}.
\enq

The system $Q$ undergoes a completely positive trace-preserving transformation
or quantum operation $\mycal{E}$ and $R$ is assumed to be isolated and its state
remains the same. This quantum operation is also represented by
$\mycal{I}_R \otimes \mycal{E}$, where $\mycal{I}_R$
is the identity superoperator on $R$.

We add subscript ``$1$" to denote the state of the system (joint or otherwise)
after this quantum operation. So the state of the joint system is denoted by
$\rho^{R_1Q_1}$. Note that
$\rho^{Q_1} = \mycal{E}(\rho)$ and $\rho^{R_1} = \rho^R$.

The entanglement fidelity is defined by Schumacher \cite{schumacher1996} as
\beq
F(\rho,\mycal{E}) = \bra{\psi^{RQ}} \rho^{R_1Q_1} \ket{\psi^{RQ} }
\enq
and the entropy exchange as
\beq
S(\rho,\mycal{E}) = S(\rho^{R_1Q_1}),
\enq
where $S(\rho^{R_1Q_1})$ is the von-Neumann entropy of $\rho^{R_1Q_1}$.
The QFI upper bounds $S(\rho,\mycal{E})$ by a function of the entanglement fidelity as
\cite{schumacher1996}
\beq
\label{qfi}
S(\rho,\mycal{E}) \leq H(F(\rho,\mycal{E})) + (1-F(\rho,\mycal{E})) \ln(d^2-1).
\enq
More details on the QFI can be found in Ref. \cite{schumacher1996},
p. 563 in Ref. \cite{nielsen-chuang}, p. 222 in Ref. \cite{hayashi}.

Generalization of the CFI
was provided by Han and Verd\'{u} \cite{han-verdu1994}, where various
lower bounds to the mutual information are given.

In this paper, we give extensions of the QFI and give various Fano-like upper bounds
to $S(\rho,\mycal{E}) $. One of the bounds that we derive for any probability vector
$\mbm{\gamma}=[\gamma_1 \cdots \gamma_d]$ is
\beqa
\label{newqfi}
S(\rho,\mycal{E}) & \leq & H(F(\rho,\mycal{E})) +
\ln\left( \sum_{i=1}^d \lambda_i
\gamma_i \right) + (1-F(\rho,\mycal{E})) ~ \ln\left( \frac{d}{\sum_{i=1}^d  \lambda_i
\gamma_i} -1 \right) 
- \sum_{k=1}^d \lambda_k \ln(\gamma_k), ~~~~~~~
\enqa
where using Eq. (\ref{eqrho}),
$\lambda_i$, $i=1,...,d$, are the eigenvalues of $\rho$.
It is easy to see that Eq. (\ref{qfi}) is a special case of Eq. (\ref{newqfi}) by substituting
$\gamma_i = 1/d$, $i=1,...,d$.
Our approach also gives an alternate derivation of the QFI.

\section{Extensions of the Quantum Fano inequality}

Let $R_2$, $Q_2$ be two ancilla quantum systems, possibly entangled,
described by $\mycal{H}_Q$. The joint system $R_2Q_2$ is described by
$\mycal{H}_{RQ} = \mycal{H}_Q \otimes \mycal{H}_Q$, and let
$\{ \ket{k^{RQ}}\}$ be an orthonormal basis for $\mycal{H}_{RQ}$, and
we define a set of projectors as
\beq
P_k = \ket{k^{RQ}} \bra{k^{RQ}}, \ \ \ \ \sum_{k=1}^{d^2} P_k = I^{RQ},
\enq
where we have chosen
\beq
\ket{1^{RQ}} = \ket{\psi^{RQ}},
\enq
and $I^{RQ}$ is the $d^2 \times d^2$ identity matrix.
Then
\beqa
S(\rho,\mycal{E}) & = & S(\rho^{R_1Q_1}) \\
& = & -S(\rho^{R_1 Q_1} || \rho^{R_2 Q_2})
- \tr(\rho^{R_1 Q_1} \ln(\rho^{R_2 Q_2})) \\
\label{temp1}
& \leq & -S\left( \sum_{k=1}^{d^2} P_k \rho^{R_1 Q_1} P_k \Big| \Big|
\sum_{k=1}^{d^2} P_k \rho^{R_2 Q_2} P_k \right)
- \tr(\rho^{R_1 Q_1} \ln(\rho^{R_2 Q_2})) \\
\label{temp2}
& = & -D( {\mathbf p} || {\mathbf q}) - \tr(\rho^{R_1 Q_1} \ln(\rho^{R_2 Q_2})),
\enqa
where
\beq
S(\rho || \sigma) = \tr(\rho \ln(\rho)) - \tr(\rho \ln(\sigma))
\enq
is the quantum relative entropy,
in Eq. (\ref{temp1}) we have used the fact that a trace-preserving completely
positive transformation reduces the quantum relative entropy
(see Refs. \cite{lindblad1975, uhlmann1977}, p. 47 in Ref. \cite{petz}),
\beqa
{\mathbf p} & = & [p_1 \cdots p_{d^2}], \\
{\mathbf q} & = & [q_1 \cdots q_{d^2}], \\
p_k & = & \bra{k^{RQ}} \rho^{R_1 Q_1} \ket{k^{RQ}}, \\
q_k & = & \bra{k^{RQ}} \rho^{R_2 Q_2} \ket{k^{RQ}},
\enqa
and $D(\cdots || \cdots)$ is the classical relative entropy given by
\beq
D( {\mathbf p} || {\mathbf q}) = \sum_{k=1}^{d^2} p_k \ln\left( \frac{p_k}{q_k}
\right).
\enq
Let
\beq
g(p,q) = D\left\{ [p, (1-p)] ~~ \big|\big|~~ [q, (1-q)]\right\}.
\enq
Then
\beqa
D( {\mathbf p} || {\mathbf q}) - g(p_1,q_1) & = &
\sum_{k=2}^{d^2} p_k \ln\left( \frac{p_k}{q_k} \right) -
(1-p_1)\ln\left( \frac{1-p_1}{1-q_1} \right) \\
& = & \sum_{k=2}^{d^2} p_k \ln\left( \frac{p_k (1-q_1)}{q_k(1-p_1)} \right) \\
\label{temp3}
& \geq & \sum_{k=2}^{d^2} p_k \left( 1 - \frac{q_k (1-p_1)}{p_k (1-q_1)} \right)
~~~~~~~~ \\
\label{bound1}
& = & 0,
\enqa
where in Eq. (\ref{temp3}), we have used the fact that for $x > 0$, $\ln(x) \geq 1 - 1/x$,
with equality if and only if $x=1$. Hence, the equality condition for Eq. (\ref{bound1}) is
\beq
\frac{q_k}{p_k} = \frac{1-q_1}{1-p_1}, ~~ k=2, \cdots, d.
\enq
More general lower bounds to the classical relative entropy are given by
Blahut in Ref. \cite{blahut1976}.
Substituting Eq. (\ref{bound1}) into Eq. (\ref{temp2}), we get
\beq
\label{ineq1}
S(\rho,\mycal{E}) \leq -g\left[F(\rho,\mycal{E}),q_1\right] - \tr\left[\rho^{R_1 Q_1}
\ln(\rho^{R_2 Q_2})\right],
\enq
where we have used the fact that $p_1 = F(\rho,\mycal{E})$. There are different choices of
the $\rho^{R_2 Q_2}$ possible to give different upper bounds on $S(\rho,\mycal{E})$.
We consider a few such choices below.

\section{Special Cases}

Let
\beq
\label{r2q2}
\rho^{R_2Q_2} = \sum_{k=1}^d \gamma_k \ket{k^R} \bra{k^R} \otimes \rho^{Q_2},
\enq
where
$\mbm{\gamma} = [\gamma_1 \cdots \gamma_d]$ is a probability vector,
and we have not yet specified the state $\rho^{Q_2}$. This choice yields
\beqa
q_1 & = & \sum_{i,j,k=1}^d \sqrt{\lambda_i \lambda_j} \gamma_k
\bra{i^R}\bra{i^Q} ~ \left(  \ket{k^R}\bra{k^R} \otimes \rho^{Q_2} \right) ~
\ket{j^R} \ket{j^Q} ~~~~~~~~ \\
& = & \sum_{i,j,k=1}^d \sqrt{\lambda_i \lambda_j} \gamma_k \delta_{i,k} \delta_{k,j}
\bra{i^Q} \rho^{Q_2} \ket{j^Q} \\
& = & \sum_{k=1}^d \gamma_k \lambda_k \bra{k^Q} \rho^{Q_2} \ket{k^Q},
\enqa
where $\delta_{i,k} = 1$ if $i=k$ and is zero otherwise.
Using Eq. (\ref{ineq1}), we get
\beqa
\label{ineq2}
S(\rho,\mycal{E}) & \leq & -g(F(\rho,\mycal{E}),q_1) -
\sum_{k=1}^d \lambda_k \ln(\gamma_k)
- \tr\left(\mycal{E}(\rho) \ln(\rho^{Q_2}) \right), ~~~~
\enqa
where we have used $\rho^{Q_1} = \mycal{E}(\rho)$.
Again, different choices of $\rho^{Q_2}$ are possible. Let us consider
\beq
\rho^{Q_2} = \sum_{k=1}^d \xi_k \ket{k^Q}\bra{k^Q},
\enq
where $\mbm{\xi} = [\xi_1 \cdots \xi_d]$ is a probability vector.
With this choice and noting that
\beqa
-\tr\left[\mycal{E}(\rho) \ln(\rho^{Q_2}) \right] & = & -\sum_k \ln(\xi_k)
\bra{k_Q} \mycal{E}(\rho) \ket{k_Q} \\
& \leq & -\ln(\min_i \{\xi_i\}).
\enqa
Eq. (\ref{ineq2}) reduces to
\beqa
S(\rho,\mycal{E}) & \leq & -g\left(F(\rho,\mycal{E}),\sum_{k=1}^d \lambda_k
\gamma_k \xi_k\right)
- \sum_{k=1}^d \lambda_k \ln(\gamma_k) - \ln(\min_i \{\xi_i\}) \\
& = & H(F(\rho,\mycal{E})) + \ln\left( \frac{\sum_{i=1}^d \lambda_i
\gamma_i \xi_i}{\min_i\{\xi_i\}} \right)
+ (1-F(\rho,\mycal{E})) ~ \ln\left( \frac{1}{\sum_{i=1}^d  \lambda_i
\gamma_i \xi_i} -1 \right) \nonumber \\
\label{ineq3}
& & ~~~~ - \sum_{k=1}^d \lambda_k \ln(\gamma_k),
\enqa
where $H(\cdots)$ is given by Eq. (\ref{binent}).

The QFI follows as a special case by substituting $\gamma_k = \xi_k = 1/d$,
$k=1,...,d$.
Note that the above inequality holds for any probability vectors $\mbm{\gamma}$ and
$\mbm{\xi}$.
We get the following simpler bound than Eq. (\ref{ineq3}) by
choosing $\xi_k = 1/d$, $k=1,...,d$,
\beqa
\label{ineq4}
S(\rho,\mycal{E}) & \leq & H(F(\rho,\mycal{E})) +
\ln\left( \sum_{i=1}^d \lambda_i
\gamma_i \right) 
+ (1-F(\rho,\mycal{E})) ~ \ln\left( \frac{d}{\sum_{i=1}^d  \lambda_i
\gamma_i} -1 \right)
- \sum_{k=1}^d \lambda_k \ln(\gamma_k). ~~~~~~~
\enqa

Eqs. (\ref{ineq1}), (\ref{ineq2}), (\ref{ineq3}), and (\ref{ineq4}) are various Fano-like
bounds that can be made tighter by appropriately choosing $\rho^{R_2Q_2}$,
$\{\mbm{\gamma}, \rho^{Q_2}\}$, $\{\mbm{\gamma}, \mbm{\xi}\}$, and
$\mbm{\gamma}$ respectively.

It might seem that one could get away from the dependence of the bounds on
$\mbm{\lambda}$ by making the following choice of $\rho^{R_2Q_2}$, which
is different from Eq. (\ref{r2q2}). Let $\beta_k$, $k= 1,...,d^2$, be the
eigenvalues of $\rho^{R_2Q_2}$ and $\ket{\psi^{RQ}}$
be one of the eigenvectors of $\rho^{R_2Q_2}$. Let $\beta_{\max} = \max_k \beta_k$,
$\beta_{\min} = \min_k \beta_k$. Since the maximum of
$g(F,x)$, $x \in [\beta_{\min},\beta_{\max}]$ occurs at the end-points, hence
to make the bound tighter,
one could choose the eigenvalue corresponding to the eigenvector $\ket{\psi^{RQ}}$
as either $\beta_{\min}$ or $\beta_{\max}$.
The bound in Eq. (\ref{ineq1}) can be simplified to
\beq
\label{temp4}
S(\rho,\mycal{E}) \leq - g(F(\rho,\mycal{E}),q_1) - \ln(\beta_{\min}),
\enq
where $q_1 = \beta_{\max}$ or $q_1 = \beta_{\min}$.
Suppose $q_1 = \beta_{\max}$, then to tighten the bound, one could
choose $\beta_{\min}$ as large as possible, or
\beq
\beta_{\min} = \frac{1-\beta_{\max}}{d^2-1}.
\enq
Substituting in Eq. (\ref{temp4}), we get
\beqa
\label{temp5}
S(\rho,\mycal{E}) & \leq & H(F(\rho,\mycal{E})) - F(\rho,\mycal{E})
\ln\left( \frac{1}{\beta_{\max}} -1 \right) + \ln(d^2-1).
\enqa
We get the tightest bound by choosing minimum value of $\beta_{\max}$ given
by $\beta_{\max} = 1/d^2$, which reduces Eq. (\ref{temp5}) to the QFI.

If $q_1 = \beta_{\min}$, then Eq. (\ref{temp4}) reduces to
\beqa
\label{temp6}
S(\rho,\mycal{E}) & \leq & H(F(\rho,\mycal{E})) + \left[1- F(\rho,\mycal{E}) \right]
\ln\left( \frac{1}{\beta_{\min}} -1 \right).
\enqa
We get the tightest bound by choosing maximum value of $\beta_{\min}$ given
by $\beta_{\min}=1/d^2$, which reduces Eq. (\ref{temp6}) to the QFI.
Hence, this choice of $\rho^{R_2Q_2}$ offers no improvement over the QFI.

\section{An Example}

We compute the QFI and the proposed inequality in Eq. (\ref{ineq4})
for the depolarizing channel for a single qubit ($d=2$) given by
\beq
\mycal{E}(\rho) = \left(1-\frac{3p}{4}\right)\rho +
\frac{p}{4} (X\rho X + Y\rho Y + Z\rho Z),
\enq
where $X,Y,Z$ are Pauli matrices. Let
\beq
\rho = U
\left(
\begin{array}{rr}
\lambda & 0 \\
0 & 1-\lambda
\end{array}
\right)
U^\dagger,
\enq
where $U$ is a randomly chosen $2 \times 2$ Unitary matrix. It is easy to show that
for any choice of $U$
\beq
F(\rho,\mycal{E}) = 1 + p \left( \lambda^2 - \lambda - \frac{1}{2} \right),
\enq
\beqa
S(\rho,\mycal{E}) & = & H_S(\acute{\mbm{\lambda}}),
\enqa
where $H_S(\cdots)$ is the Shannon entropy, $\acute{\mbm{\lambda}} = \left[
p\lambda/2, (1-\lambda)p/2, -p/4 +1/2 +\theta/4, -p/4 +1/2 -\theta/4 \right]$,
and
\beq
\theta = \sqrt{p^2 +12p^2\lambda(1-\lambda) + 4(1-p) -16p\lambda(1-\lambda)}.
\enq

\begin{figure}[ht]
\centering{
\resizebox{12.5cm}{!}{\includegraphics[angle=-90]{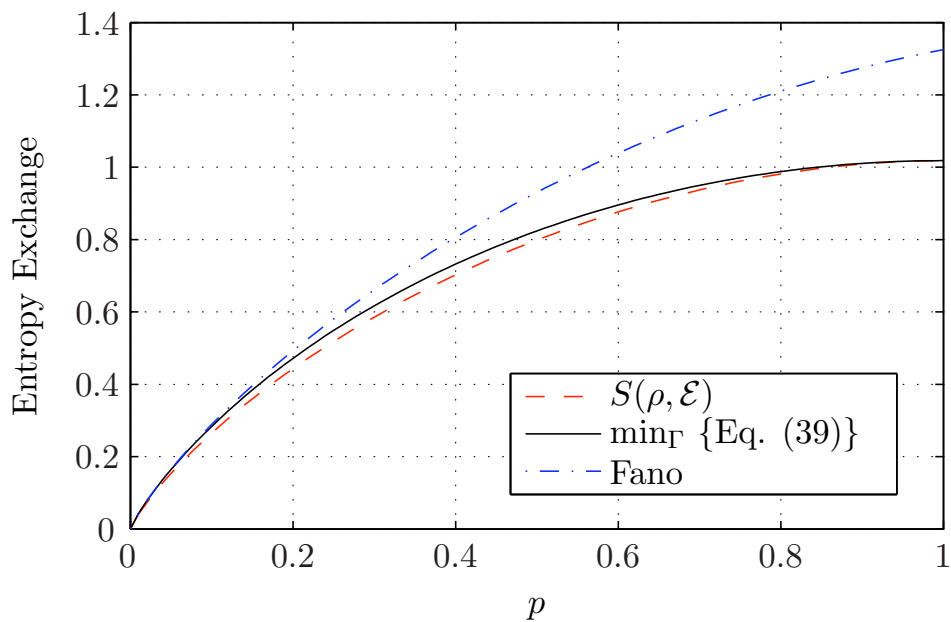}}
}
\caption{\label{fig1} Plots of $S(\rho,\mycal{E})$, the tightest bound from
Eq. (\ref{ineq4}), and the QFI.}
\end{figure}

\begin{figure}[ht]
\centering{
\resizebox{12.5cm}{!}{\includegraphics[angle=-90]{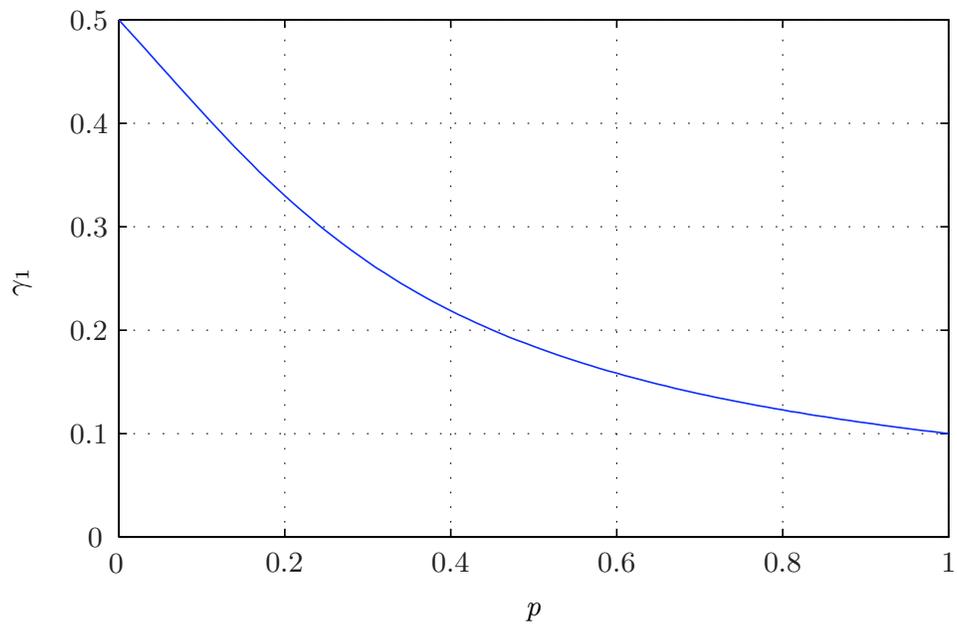}}
}
\caption{\label{fig2} $\gamma_1$ that gives the tightest bound in Eq. (\ref{ineq4}).}
\end{figure}

In Fig. \ref{fig1}, we compare $S(\rho,\mycal{E})$ with the QFI and
the inequality in Eq. (\ref{ineq4}) numerically optimized over $\mbm{\gamma}$ to
give the tightest bound for $\lambda = 0.1$. The figure shows
that the latter bound is tighter than the QFI.
In Fig. \ref{fig2}, we plot the numerically computed
value of $\gamma_1$ that gives the tightest bound in Eq. (\ref{ineq4}). The QFI
corresponds to $\gamma_1=1/d=0.5$.

\newpage

\bibliographystyle{unsrt}

\end{document}